\documentclass[aps,prl,twocolumn,showpacs,superscriptaddress,groupedaddress]{revtex4}  

\usepackage{graphicx}  
\usepackage{dcolumn}   
\usepackage{bm}        
\usepackage{amssymb}   

\usepackage{xspace}
\usepackage{amsmath}
\usepackage{abbrevs}
\usepackage[]{units}	

\usepackage{float} 
\usepackage{wrapfig} 

\newcommand{\ket}[1]{| #1 \rangle}

\newcommand{\gtwo}[1]{g^{(2)}_{\mathrm{#1}}(0)}

\newabbrev\SPDC{spontaneous parametric downconversion (SPDC)\xspace}[SPDC]
\newabbrev\APD{avalanche photodiode (APD)\xspace}[APD]
\newabbrev\TBP{time-bandwidth product (TBP)\xspace}[TBP]
\newabbrev\FWM{four-wave mixing (FWM)\xspace}[FWM]
\newabbrev\SNR{signal-to-noise ratio (SNR)\xspace}[SNR]

\makeatletter
\renewcommand\maybe@space@{%
  \maybe@ictrue 
  \expandafter   \@tfor
    \expandafter \reserved@a
    \expandafter :%
    \expandafter =%
                 \nospacelist
                 \do \t@st@ic
  \ifmaybe@ic 
    \space
  \fi
}
\makeatother

\begin{document}



\title{Storage and retrieval of ultrafast single photons using a\\ room-temperature diamond quantum memory}
\author{Duncan~G.~England} \affiliation{National Research Council of Canada, 100 Sussex Drive, Ottawa, Ontario, K1A 0R6, Canada}

\author{Kent~A.G.~Fisher}\affiliation{Institute for Quantum Computing and Department of Physics \& Astronomy, University of Waterloo, 200 University Avenue West, Waterloo, Ontario N2L 3G1, Canada}

\author{Jean-Philippe~W.~MacLean}\affiliation{Institute for Quantum Computing and Department of Physics \& Astronomy, University of Waterloo, 200 University Avenue West, Waterloo, Ontario N2L 3G1, Canada}

\author{Philip~J.~Bustard} \affiliation{National Research Council of Canada, 100 Sussex Drive, Ottawa, Ontario, K1A 0R6, Canada}

\author{Rune Lausten} \affiliation{National Research Council of Canada, 100 Sussex Drive, Ottawa, Ontario, K1A 0R6, Canada}

\author{Kevin~J.~Resch}\affiliation{Institute for Quantum Computing and Department of Physics \& Astronomy, University of Waterloo, 200 University Avenue West, Waterloo, Ontario N2L 3G1, Canada}

\author{Benjamin J. Sussman} \affiliation{National Research Council of Canada, 100 Sussex Drive, Ottawa, Ontario, K1A 0R6, Canada}

\date{\today}

\begin{abstract}

We report the storage and retrieval of single photons, via a quantum memory, in the optical phonons of room-temperature bulk diamond. The THz-bandwidth heralded photons are generated by spontaneous parametric downconversion and mapped to phonons via a Raman transition, stored for a variable delay, and released on demand. The second-order correlation of the memory output is $\gtwo{}=0.65\pm0.07$, demonstrating preservation of non-classical photon statistics throughout storage and retrieval. The memory is low-noise, high-speed and broadly tunable; it therefore promises to be a versatile light-matter interface for local quantum processing applications.

\end{abstract}

\maketitle

Single photons are challenging to create, manipulate and measure, yet are essential for a diverse range of quantum technologies, including cryptography~\cite{Bennett1984}, enhanced measurement~\cite{Giovannetti2004}, and information processing~\cite{Knill2001}. Quantum memories, which act as buffers for photonic states, are a key enabling component for these future technologies~\cite{Bussieres2013}. They allow repeat-until-success strategies to counteract the intrinsically probabilistic nature of quantum mechanics, thereby providing scalable quantum technologies. An ideal quantum memory would store a single photon, maintain the quantum state encoded in the photon, and release it, on-demand, as a faithful recreation of the input. Efforts to implement optical quantum memories have used a number of platforms including single atoms in a cavity~\cite{Specht2011}, ultracold atoms~\cite{Choi2008}, atomic vapours ~\cite{Reim2010}, molecular gases~\cite{Bustard2013} and rare-earth doped crystals~\cite{Hedges2010}. 

The potential for quantum storage in optical memories is often investigated using laser pulses attenuated to the single-photon level~\cite{Riedmatten2008}. However, the transition between storing weak coherent states and true single photons produces two significant obstacles. Firstly, to achieve high efficiency, most memories must operate near resonance with a dipole transition, typically limiting storage bandwidths to $\sim$GHz or below~\cite{Reim2010,Saglamyurek2011}. Single photon sources compatible with such devices require careful engineering to match the frequency and bandwidth of the photons to that of the memory~\cite{Choi2008,Saglamyurek2011,Chaneliere2005,Zhang2011,Clausen2011,Rielander2014}. Secondly,  the intense read and write beams used to mediate storage and retrieval may introduce noise which obscures the quantum properties of the signal~\cite{Michelberger2014}. Where non-classical memory operation has been demonstrated, laser-cooled~\cite{Choi2008,Chaneliere2005,Zhang2011} or cryogenic~\cite{Saglamyurek2011,Clausen2011,Rielander2014} substrates are often required to reduce noise.

In this letter, we demonstrate the storage and retrieval of broadband single photons using a room-temperature solid-state quantum memory. The THz-bandwidth heralded single photons are created by \SPDC and are stored, via an off-resonant Raman transition, in the optical phonon modes of a room-temperature bulk diamond. As the Raman interaction occurs far from any optical resonances, the memory can operate at a range of visible and near-infrared wavelengths. The bandwidth of the memory is limited only by the 40\,THz splitting between the ground and storage states~\cite{Solin1970}. This broad bandwidth and large tuning range makes the memory compatible with ultrafast \SPDC photon sources~\cite{Mosley2008}. Furthermore, the memory exhibits a quantum-level noise floor, even at room temperature~\cite{England2013}.

\begin{figure*} 
\center{\includegraphics[width=0.8\linewidth]{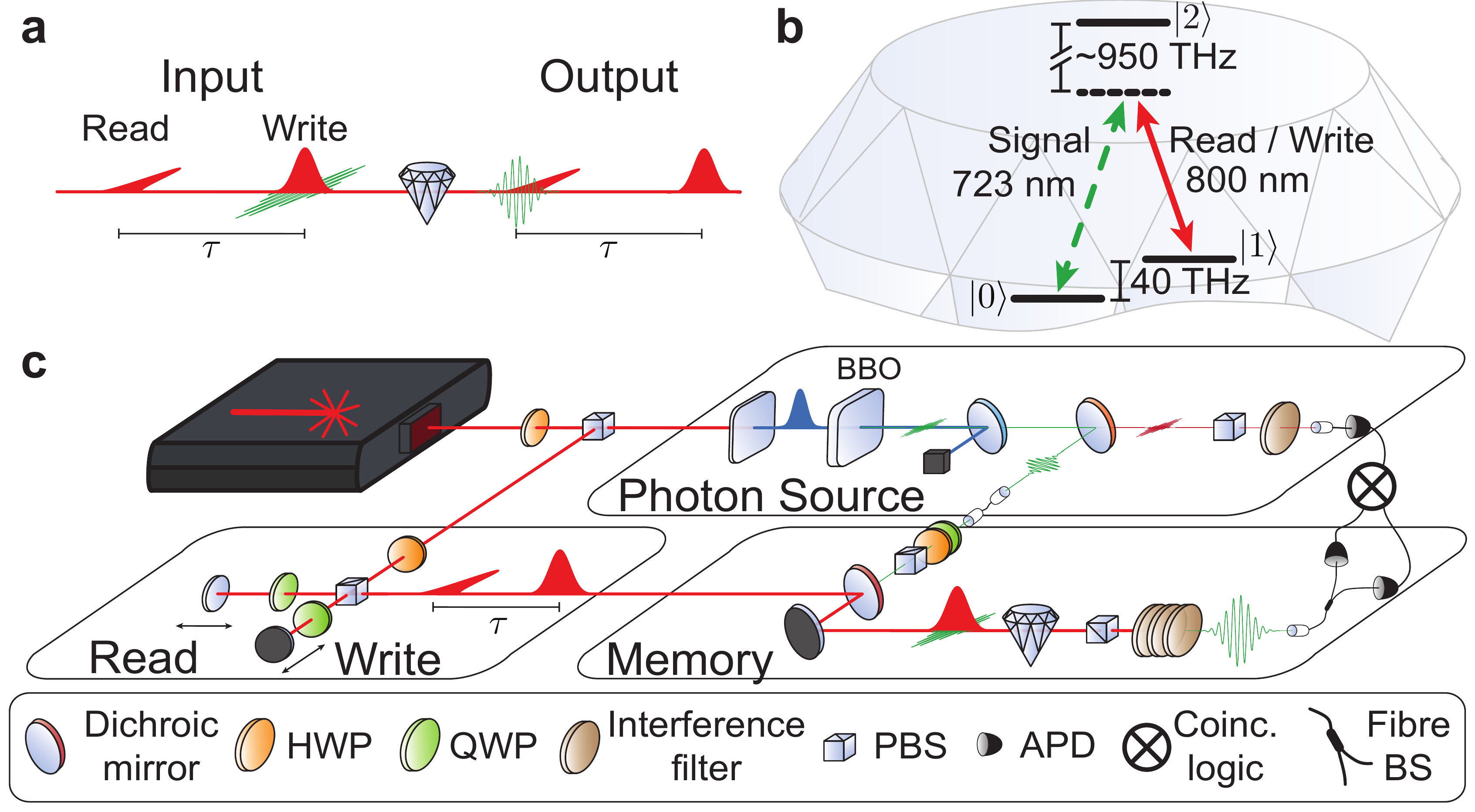}}
\caption{Experimental concept, energy level diagram, and setup. {\bf a:} The memory protocol. A horizontally (H) polarized single photon (green, 723\,nm) is written into the quantum memory with a vertically (V) polarized \emph{write} pulse (red, 800\,nm). After a delay $\tau$ an H-polarized \emph{read} pulse recalls a V-polarized photon. {\bf b:} Energy levels in the memory. The ground state $\ket{0}$ and storage state $\ket{1}$ correspond to the crystal ground state and an optical phonon respectively. The signal photon and the read/write pulses are in two-photon resonance with the optical phonon (40\,THz) and are far detuned from the conduction band $\ket{2}$. {\bf c:} The experimental setup. The laser output is split to pump the photon source and to produce the orthogonally-polarized read and write beams. The photons are produced in pairs with one (signal) at 723\,nm and the other (herald) at 895\,nm. The signal photon is stored in, and recalled from, the quantum memory. The herald and signal photons are detected using APDs and correlations between them are measured using a coincidence logic unit.}
\label{fig:Setup}
\end{figure*}

The use of an ultrafast laser oscillator is essential to this device. By applying pulses of $\sim$200\,fs duration, we address the memory on timescales that are short compared to its decoherence time, allowing multiple operational time-bins during storage. Furthermore, the fast repetition rate of the laser provides single photons at a high rate, thereby returning a good signal-to-noise ratio in photon-counting experiments, on a reasonable timescale. The combination of ultrafast lasers and diamond phononics provides a convenient high-speed system in which to build and test quantum devices. 

The memory utilized a high-purity, low-birefringence synthetic diamond manufactured by Element Six Ltd. The diamond, which is 2.3\,mm thick, was grown by chemical vapour deposition and is cut along the $\left<100\right>$ face of the diamond lattice. The relevant energy levels can be described by a $\Lambda$-level system consisting of the crystal ground state $\ket{0}$, an optical phonon acting as the storage state $\ket{1}$ and an off-resonant intermediate state $\ket{2}$ representing the conduction band. Single photons are stored to, and retrieved from, the optical phonon by strong \emph{write} and \emph{read} pulses via an off-resonant Raman interaction~\cite{Nunn2007,Reim2010}. The photons and read/write pulses are in two-photon resonance with the optical phonon energy (see Fig.~\ref{fig:Setup}b). In the $\left<100\right>$ configuration, the Raman interaction couples fields of orthogonal polarization~\cite{Hayes1978} such that the H-polarized input photons are stored by a V-polarized write pulse and V-polarized output photons are retrieved using a H-polarized read pulse.

The high carrier frequency of the optical phonon (40\,THz~\cite{Solin1970}) and a large detuning from the conduction band ($\sim950$\,THz) are the key features allowing storage of THz-bandwidth photons. These features also provide an intrinsically low noise floor: the large detuning from optical resonance eliminates fluorescence noise, and the high energy of the optical phonon results in low thermal phonon population at room temperature. Four-wave mixing noise, which is a pervasive problem in many $\Lambda$-level systems~\cite{Lauk2013,Michelberger2014}, is suppressed in diamond due to the large splitting and high optical dispersion~\cite{England2013}. Following excitation, the optical phonons decay into a pair of acoustic phonons with a characteristic timescale of 3.5\,ps~\cite{Lee2012}, which sets the storage lifetime of the memory. While this lifetime is prohibitively short for some applications, the advantage of the rapid acoustic decay is that it returns the crystal lattice to the ground state, resetting the memory such that it is ready to store the next photon. This sub-nanosecond reset time permits GHz repetition rates in the diamond phonon system.

The master laser for the experiment is a mode-locked Ti:sapphire laser producing pulses of 190\,fs duration at a repetition rate of 80\,MHz, central wavelength of 800\,nm and a pulse energy of 28\,nJ. The laser beam is split between the photon source and the memory with 12.5\,nJ used to generate the orthogonally-polarized read and write pulses (read/write panel, Fig.~\ref{fig:Setup}c). The remaining energy for the photon source is frequency-doubled in a 1\,mm $\beta$-barium borate (BBO) crystal to produce pulses at 400\,nm (pulse energy 2.4\,nJ). In a second 1\,mm BBO crystal, angle tuned to phase-match type-I non-degenerate \SPDC, the pump field at 400\,nm produces horizontally-polarized photon pairs at 723\,nm (signal) and 895\,nm (herald). The photon pairs are emitted collinearly from the BBO and, after the remaining 400\,nm pump light has been removed by interference filters, the signal and herald photons are spatially separated by an 801\,nm long-pass dichroic mirror.  The herald photons pass through a polarizing beam splitter (PBS) and a 5\,nm bandwidth interference filter before being coupled into a single-mode fiber and detected on an \APD.  The signal photons are coupled into a 7\,cm long single-mode fiber for spatial filtering before being directed to the memory. 

The horizontally-polarized signal photon is spatially and temporally overlapped on a dichroic mirror with the vertically-polarized write pulse and focused into the diamond for storage by a 6\,cm focal length achromatic lens. After a time delay $\tau$ the horizontally-polarized read pulse arrives at the memory and the photon is re-emitted, this time with vertical polarization. We can thus distinguish between the input and the output states of the memory by their polarization (Fig.~\ref{fig:Setup}a). Following the memory, the signal photons are spectrally filtered from the read/write pulses and coupled into a single-mode fiber. By rotating the polarization basis we collect either the memory output or the unabsorbed memory input (memory panel, Fig.~\ref{fig:Setup}c). The photons are detected by an \APD and correlations in photon detection events are measured using coincidence counting logic.

\begin{figure} 
\center{\includegraphics[width=0.9\linewidth]{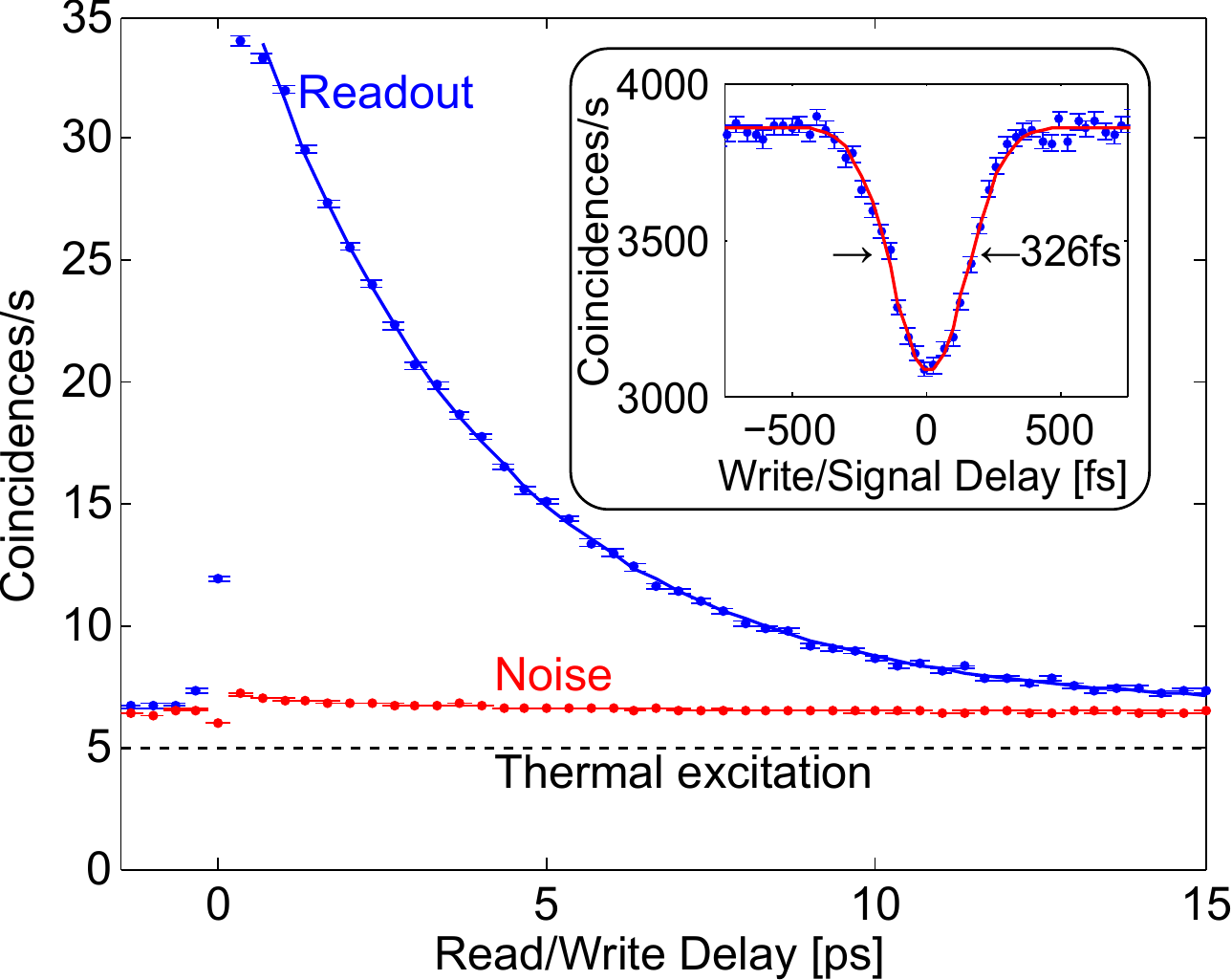}}
\caption{Measured coincidences between the signal and herald photons as a function of read-write delay (blue bars). An exponential decay of half-life 3.5\,ps (solid blue fit) is characteristic of the optical phonon lifetime. The background noise (red bars) shows a SNR of 3.8:1 for single photon retrieval. The estimated noise due to thermal phonon population is shown by the dotted line. {\bf Inset:} Transmission through the diamond in the presence of the write pulse, showing memory absorption. The profile width of 326\,fs (solid red fit) indicates the large bandwidth of the photons stored in the memory. All error bars are from poissonian counting statistics. Signal and herald detection events are defined as coincident if the time delay between them falls within a 1\,ns window.}
\label{fig:Readout}
\end{figure}

Storage of the signal photons is demonstrated by scanning the delay of the write pulse with respect to the signal photon. With 12.5\,nJ in the write pulse, a 20\% reduction in signal-herald coincidences at zero delay indicates that signal photons are being written to the memory (inset, Fig.~\ref{fig:Readout}). The full width at half maximum of the absorption profile is $w_\mathrm{a} = 326$\,fs. Deconvolving this width with that of the write pulse ($w_\mathrm{w} = 190$\,fs) using the expression $w_\mathrm{a}^2 = w_\mathrm{w}^2 + w_\gamma^2$ returns an estimated photon duration of $w_\gamma = 260$\,fs, assuming transform limited pulses with Gaussian spectra. 

Readout of the signal photons is observed by rotating the polarization filter to measure the vertically-polarized output of the memory. With 6.25\,nJ in each pulse, we scan the delay between the write and read pulses; the sharp step in signal-herald coincidences at zero delay indicates that signal photons are being retrieved from the memory (Fig. \ref{fig:Readout}). The maximum total memory efficiency is $\eta_\mathrm{t}=0.9\%$ and the write efficiency is $\eta_\mathrm{w}=9\%$ from which we extract a read efficiency of $\eta_\mathrm{r}=\eta_\mathrm{w}/\eta_\mathrm{t}=10\%$. The exponential decay in read efficiency has a half-life of 3.5\,ps which is characteristic of the optical phonon lifetime~\cite{Lee2012}. We have therefore demonstrated that the memory stores a single photon for over 13 times its duration.

By blocking the input signal photons we can measure the background noise of the memory (see Fig.~\ref{fig:Readout}), the maximum \SNR is 3.8:1; it is important to note that this is a raw measurement and no background subtraction has been performed. This noise has two origins: spontaneous anti-Stokes scattering from thermally-excited phonons and spontaneous \FWM. The \FWM noise process is intrinsic to the memory and cannot be completely eliminated, however in a dispersive material such as diamond it is strongly suppressed due to phase-matching conditions~\cite{England2013}. From Boltzmann statistics we calculate that around 5 coincidence counts per second can be attributed to thermal noise (dashed line in Fig.~\ref{fig:Readout}). The photon heralding efficiency at the memory is 16\% meaning that a single photon is present at the memory in only 16\% of the heralded experiments; despite this, we still observe a high contrast between signal and noise. With an ideal heralding efficiency, and no thermal noise, the possible \SNR in this memory is in excess of 70:1.

\begin{figure} 
\center{\includegraphics[width=0.9\linewidth]{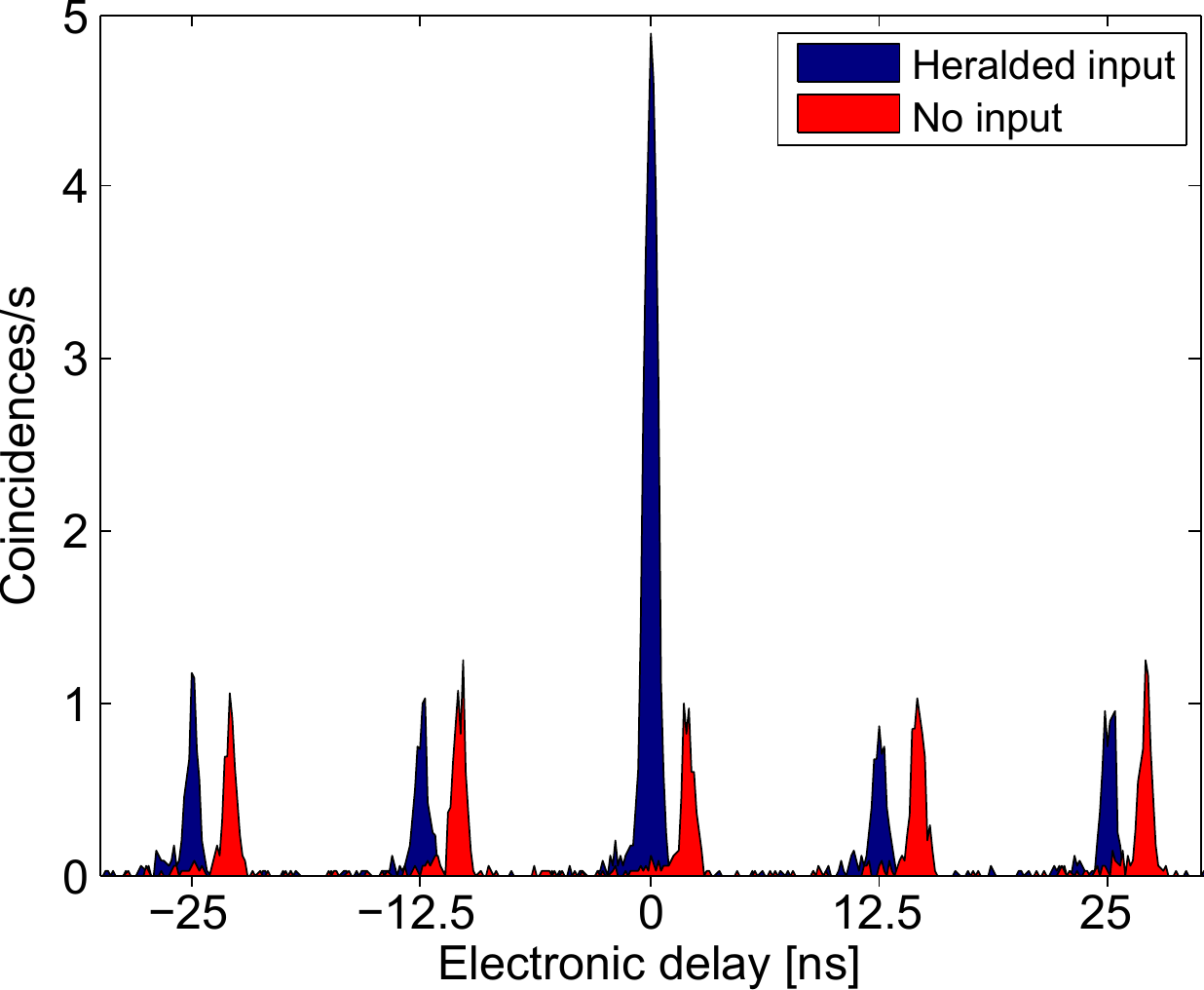}}
\caption{Detection coincidences between the herald photon and the signal photon retrieved from the memory as a function of electronic delay (blue). A peak of five times the accidental rate at zero delay demonstrates strong correlations between the herald and signal photons after readout from the memory. The memory noise (red, offset by 2\,ns for clarity) shows no increase at zero delay. Signal and herald detection events are defined as coincident if the time delay between them falls within a 156\,ps window. The width of the peaks is due to the timing jitter ($\sim$500\,ps) of the APDs.}
\label{fig:Histogram}
\end{figure}

Correlations between the memory output and the herald photon can be seen in their coincidence statistics, as shown in Fig.~\ref{fig:Histogram}. The coincidence rate as a function of the electronic delay between the memory output and the herald photon shows periodic peaks due to accidental coincidences between the memory noise and the herald photon; the time period of these peaks is 12.5\,ns corresponding to the repetition period of the laser oscillator. The largest peak, at zero delay, indicates retrieval of signal photons which have been written to the memory. These coincidence rates exceed the accidental rate by a factor of five, a clear indication that the non-classical correlations between herald and signal are maintained during storage and retrieval from the quantum memory.

A stringent test for non-classical photon statistics is to measure the second-order correlation function $\gtwo{}$~\cite{Glauber1963}. Using the Hanbury Brown and Twiss configuration~\cite{HanburyBrown1956}, the input light field is partitioned between two detectors using a 50:50 fiber beamsplitter, as shown in Fig.~\ref{fig:Setup}c.
The triggered $\gtwo{}$ function of the heralded \SPDC source is calculated as~\cite{Grangier1986}:
\begin{equation}\label{eq:g2}
\gtwo{} = \frac{N_\mathrm{h,1,2}N_\mathrm{h}}{N_\mathrm{h,1}N_\mathrm{h,2}},
\end{equation}
where $N_\mathrm{h}$ is the number of herald photons detected in a given time window, $N_\mathrm{h,1}$ ($N_\mathrm{h,2}$) is the number of two-fold coincident detections between the herald and output port 1 (2) of the beam splitter and $N_\mathrm{h,1,2}$ is the number of three-fold coincidences between the herald and both ports of the beamsplitter. A correlation function of $\gtwo{}<1$ is a direct measure of sub-poissonian statistics which cannot be explained classically, and is evidence of single photons. 

\begin{figure} 
\center{\includegraphics[width=0.95\linewidth]{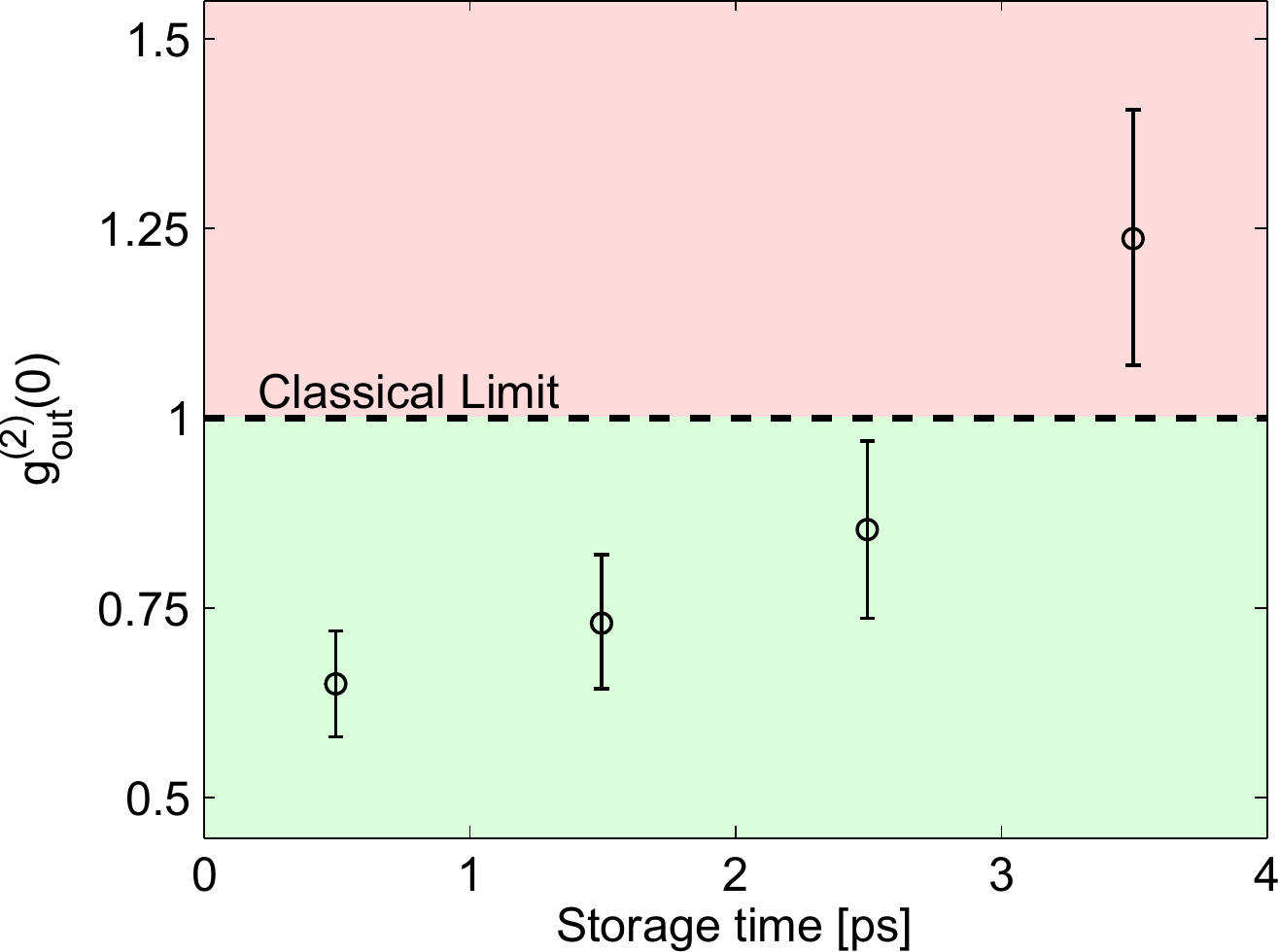}}
\caption{The heralded second-order correlation function of the memory output as a function of storage time. Values below the classical limit of $\gtwo{out} = 1$ demonstrate the quantum characteristics of the output field. Non-classical statistics are observed for storage times up to $\sim$3\,ps. Error bars are from Poissonian counting statistics.}
\label{fig:g2}
\end{figure}

At the memory input, we measure $\gtwo{in} = 0.04\pm0.01$; after the memory the correlation function $\gtwo{out}$ depends on the storage time, as shown in Fig.~\ref{fig:g2}. When the storage time is 0.5\,ps, we measure $\gtwo{out} = 0.65\pm0.07$, five standard deviations below the classical limit of 1.  The measured $\gtwo{out}$ function increases with increasing storage time as the noise comprises a larger fraction of the measured counts, however the memory output maintains non-classical statistics for $>$2.5\,ps. A three-fold coincident detection occurs, on average, once every 60\,billion laser pulses. However, because of the 80\,MHz clock rate of the experiment we are able to achieve sufficient count-rates to measure $\gtwo{}$; this illustrates the importance of high memory repetition rate for multi-photon coincidence measurements.

In conclusion, we have demonstrated a THz-bandwidth quantum memory for light using the optical phonon modes of a room-temperature diamond. The unique features of the memory enable storage of single  photons produced by ultrafast spontaneous parametric downconversion --- the most widespread source of single and entangled photons. We verified that the quantum nature of the photons is preserved during the memory interaction by demonstrating non-classical photon statistics in the memory output. The heralded second-order correlation function of the memory output was $\gtwo{out} = 0.65\pm0.07$ which is five standard deviations below the classical limit. The device requires no cooling or optical preparation before storage and is a few millimetres in size; diamond is therefore a robust, convenient and high-speed testbed system in which to evaluate operational memory parameters, study the effects of noise, and develop quantum protocols. Beyond quantum memory, this system can also be applied to generate quantum random numbers~\cite{Bustard2011} and macroscopic entanglement~\cite{Lee2011a}. In the future, its ease of use may benefit other applications such as quantum frequency conversion~\cite{McGuinness2010}, memory-enhanced optical non-linearites~\cite{Hosseini2012}, or programmable linear-optical components~\cite{Reim2012}.

The authors thank Matthew Markham and Alastair Stacey of Element Six Ltd. for the diamond sample. They also thank Paul Hockett and Josh Nunn for useful discussions, and John Donohue for writing data acquisition software. Doug Moffatt and Denis Guay provided invaluable technical assistance. This work was supported by the Natural Sciences and Engineering Research Council of Canada, Canada Research Chairs, Canada Foundation for Innovation, Ontario Centres of Excellence, and Ontario Ministry of Research and Innovation Early Researcher Award.

\bibliography{DGE}{}
\bibliographystyle{unsrt}
\end{document}